\documentclass[12pt]{article}
\usepackage{amsmath,amssymb}
\usepackage{graphicx}
\usepackage[hypertex]{hyperref}
\newcommand{\bea}{\begin{eqnarray}}
\newcommand{\eea}{\end{eqnarray}}
\def\la{\label}

\numberwithin{equation}{section}

\def\ssp{\hspace{0.3mm}}
\def\({\left(}
\def\){\right)}
\renewcommand{\[}{\left[}

\def\bb#1{\mathbb{#1}}

\def\AdS#1{AdS${}_{#1}$}

\def\AdS{AdS${}_5 \times {}$S${}^5\ $}
\renewcommand{\eqref}[1]{$\({\rm \ref{#1}}\)$}

\def\tp{{\widetilde p}}

\def\cN{{\mathcal N}}
\def\cO{{\mathcal O}}

\def\sn{\,\mathrm{sn}}
\def\cn{\,\mathrm{cn}}

\def\a {\alpha}

\def\e{\epsilon}

\def\ov{\over}
\def\tH{\widetilde{{\cal E}}}

\newcommand{\alg}[1]{\mathfrak{#1}}
\newcommand{\su}{\alg{su}}

\def\psu{\alg{psu}}
\def\sl(2){\alg{sl}(2)}

\usepackage{hyperref}
\begin{document}
%\draft
%%%%%%%%%

% Front page here

%\vspace*{1cm}
\null\vskip-40pt
%\hfill {\tt hep-th/yymmnnn}
 \vskip-5pt  \hfill {\tt\footnotesize
TCD-MATH-09-14}
 \vskip-5pt \hfill {\tt\footnotesize
HMI-09-07}

\vskip 1cm \vskip0.2truecm
\begin{center}
\begin{center}
\vskip 0.8truecm {\Large\bf Temperature 
 quantization from the TBA equations}
%\vskip 0.8truecm {\Large\bf Spectrum of \AdS Superstrings \\
%\vskip 0.1cm and the AdS/CFT Correspondence }
\end{center}

\renewcommand{\thefootnote}{\fnsymbol{footnote}}

\vskip 0.9truecm
%\author{
Sergey Frolov\footnote[1]{Email: 
frolovs, rsuzuki@maths.tcd.ie}{}\footnote[2]{ Correspondent fellow at
Steklov Mathematical Institute, Moscow.} \  and Ryo Suzuki
 \\
\vskip 0.5cm

 {\it Hamilton Mathematics Institute and School of Mathematics, \\
Trinity College, Dublin 2,
Ireland}

\end{center}
\vskip 1cm \noindent\centerline{\bf Abstract} \vskip 0.2cm We analyze the Thermodynamic Bethe Ansatz equations for the
 mirror model which determine the ground state energy 
of the light-cone \AdS
superstring living on a cylinder.
The light-cone momentum of string is equal to the circumference of the cylinder,
and is identified with the inverse temperature of the mirror model.
We show that the natural requirement of the analyticity of the Y-functions leads to the quantization of the temperature of the mirror model which has never been observed in any other models.

\renewcommand{\thefootnote}{\arabic{footnote}}
\setcounter{footnote}{0}
%%%%%%%%%%%%%%%%%%%%%%%%%%%%%%%
%%%%%%%%%%%%%%%%%%%%%%%%%%%%%%%%%

%\vskip 0.5cm

\section{Introduction}
%%%%%%%%%%%%%%%%%%%%%%%%%%%%%%%%%

An effective way to analyze exact finite-size spectrum of two-dimensional integrable field theory models is provided by the
Thermodynamic Bethe Ansatz (TBA) approach originally developed for
relativistic models \cite{za}. Its application to nonrelativistic models such as 
the light-cone  \AdS superstring sigma model, for a review see \cite{AFrev}, requires studying the thermodynamic properties of a so-called mirror model obtained from the original one 
by a double Wick rotation. The inverse temperature of the mirror model is identified with the circumference of the cylinder the original one lives on. Then, the ground state energy
of the original model is related to the free energy (or for periodic fermions to Witten's index \cite{Witten82c}) of
the  mirror model. Moreover, it has been shown that
the TBA approach is also capable of accounting for the excited
states \cite{BLZe,DT}, see \cite{Fendley:1997ys}-\cite{GKV08}
for further results and different approaches.

In the \AdS case the mirror model was introduced and studied in detail in \cite{AFtba}.\footnote{The importance of the TBA approach for understanding the exact spectrum of the light-cone  \AdS superstring was stressed in \cite{AJK} where it was used to explain wrapping effects in gauge theory.}
In particular, the mirror model S-matrix was obtained from the \AdS world-sheet S-matrix 
\cite{S}-\cite{AFZzf} by means of a proper analytic continuation, and shown to be unitary.
It was then used to derive the Bethe-Yang (BY) equations for elementary
particles of the mirror theory which appeared to differ in a subtle but important way from the  BY equations for the light-cone  \AdS superstring and ${\cal N}=4$ SYM \cite{BS05}-\cite{Le}. 
The asymptotic spectrum of the mirror model was shown to consist of the elementary particles and 
$Q$-particle bound states which
comprise into the tensor product of two $4Q$-dimensional atypical
totally anti-symmetric multiplets of the centrally extended
$\su(2|2)$ algebra. This is in contrast to the light-cone string model where the bound states belong to the tensor product of two $4Q$-dimensional atypical
totally symmetric multiplets \cite{D06,B2}, and it played an important role in the computation of the four-loop anomalous dimension of the Konishi operator \cite{BJ08}.

The next step is to count all asymptotic states of the mirror model, and to determine its free energy. String hypothesis enables such a computation in practice \cite{Takahashi72}, and it  is the most important step towards realizing the TBA approach because TBA equations  are then easily derived following  a textbook route \cite{Korepin}. 

Recently, the results obtained in \cite{AFtba} were used to formulate a string hypothesis for the mirror model \cite{AF09a}. The derivation of the corresponding  TBA equations  was then performed in \cite{AF09b}-\cite{GKKV},\footnote{
All the three sets of TBA equations should yield the same answer provided the same string hypothesis has been used. Checking this however is not easy due to
 different notations and conventions
used in the papers.}
where the TBA equations were also used to analyze the existence of the associated Y-system
\cite{Zamolodchikov:1991et}-\cite{Bazhanov:1994ft}.
It was shown in  \cite{AF09b} that the Y-system for the planar AdS/CFT correspondence \cite{M}  conjectured in \cite{GKV09}  followed from the TBA equations only 
for the values of the rapidity variable $u$ from the interval $[-2,2]$. 
It appeared that for other values of $u$ one had to impose additional conditions
on the Y-functions. We will show in this paper that these conditions, however,  are not compatible with the ground state
energy solution of the TBA equations, and, therefore,  an {\it analytic Y-system}  does not exist. In contrast to relativistic models, the Y-system if it exists would be defined on an infinite genus Riemann surface, and 
it is unclear how such a Y-system could be used for analyzing the string spectrum along the lines of \cite{Fendley:1997ys}-\cite{Heg},\cite{GKV08}. 

Besides its potential application  to determining the full spectrum of the light-cone \AdS superstring, the mirror model has a new interesting feature as a two-dimensional quantum field theory. In the temporal gauge (see \cite{AFrev}) the light-cone momentum $P_+$ carried by the string is equal to the charge $J$ corresponding to one of the $U(1)$ isometries of $S^5$.  Since the 
superstring lives on a cylinder of circumference $L=P_+$,  the mirror theory temperature is equal to $T=1/J$. In quantum theory the charge $J$ is quantized and can take only integer or half-integer values. Consistency of the mirror model would then require quantization of the inverse temperature.\footnote{Even though this argument is based on the temporal gauge, the temperature quantization should happen also in an arbitrary light-cone gauge. The quantization condition, however, would take different forms in different gauges.}  Temperature (and length) quantization, should it happen, seems to be a new phenomenon never seen before in any other model.

In this paper we investigate the ground state energy of the light-cone string theory by using the TBA equations of \cite{AF09b}. In the sector with periodic fermions the ground state is BPS and its energy should not  receive quantum corrections. In this case, however, we encounter a singularity in the TBA equations, and regularize it through the chemical potential for fermions \cite{CFIV92}.
We then solve the TBA equations in the vicinity of the BPS vacuum, and  find that the inverse temperature is quantized at integer or half-integer values if one imposes a natural requirement of  the analyticity of the Y-functions on the $z$-torus.

%%%%%%%%%%%%%%%%%%%%%%%%%%%%%%%%%%%%%%
\section{Ground state energy}\label{sec:groundstate TBA}

In this section we use the TBA equations of \cite{AF09b} to 
compute the energy of the ground state. To save the space we do not list here all the equations, and do not provide explicit expressions for the kernels involved. The reader should consult the corresponding equations and definitions in  \cite{AF09b}. 

The energy of the ground state of the light-cone string theory depends on $L$ which in the mirror theory is equal to the inverse temperature, and in the string theory is identified with the total light-cone momentum $P_+=L$. In the  temporal gauge $P_+=J$ where $J$ is  the angular momentum carried by the string rotating about the equator of S$^5$.
The ground state energy also depends on the parameter $h$ that allows one to interpolate between the even-winding number sector, $h=0$, of the light-cone string theory with periodic fermions and supersymmetric vacuum, and the odd-winding number sector, $h=\pi$, with anti-periodic fermions and nonsupersymmetric vacuum. 
The energy is given by the following universal expression which has no explicit dependence on $h$
\begin{equation}
E_h (L) = - \int_{-\infty}^\infty {du \over 2\pi} \, \sum_{Q=1}^\infty {d\tp_Q \over du} \, 
\log ( 1 + Y_Q ), \qquad Y_Q \equiv e^{-\e_Q},
\label{energyL}
\end{equation} 
where $\tp^Q$ and $\e_Q$ are the momentum and pseudo-energy of a mirror $Q$-particle. 

For the supersymmetric vacuum we expect the ground state energy to vanish 
$E_{h=0}=0$. 
According to \eqref{energyL}, the condition $E_h=0$ requires $Y_Q=0$. We see that the whole set of TBA equations is solved by
\begin{equation}
Y_Q=0, \qquad Y_{+}^{(\a)}=Y_{-}^{(\a)}=1, \qquad Y_{M|vw}^{(\a)} = Y_{M|w}^{(\a)} \neq 0, \qquad e^{ih_\a}=1 \,.
\label{infinite sol}
\end{equation}
A subtle point here is that the TBA equation for $Q$-particles is singular at $Y_Q=0$. To regularize this singularity in the next subsection we consider the general case with $h\neq 0$ and take the limit $h\to 0$.

\subsection{Small $h$  expansion}

If $h$ is small any Y-function can be expanded in a series in $h$. 
The leading small $h$ behavior can be understood from the TBA equation for $Q$-particles
\begin{multline}
- \log Y_Q = L \, \tH_{Q} - \log\left(1+Y_{Q'} \right)\star K_{\sl(2)}^{Q'Q} - \log\left(1+{1\ov Y_{M|vw}^{(\a)}} \right)\star K^{MQ}_{vwx} \\
- {1\ov 2} \log{1-{e^{ih_\a} \ov Y_{-}^{(\a)}}\ov 1-{e^{ih_\a} \ov Y_{+}^{(\a)}} }\star K_Q - {1\ov 2} \log\left(1-{e^{ih_\a} \ov Y_{-}^{(\a)}} \right) \left(1-{e^{ih_\a} \ov Y_{+}^{(\a)}}\right)\star K_{yQ} \,,
\label{YforQ}
\end{multline}
where 
\bea \la{def:mirror energy}
 \tH_{Q} = 2 \, {\rm arcsinh} ( \frac{\sqrt{Q^2 + \tp^2}}{2 \ssp g} )\,,
 \eea 
 is the energy of a $Q$-particle,
$h_\a=(-1)^\a h$, the summation over $\alpha=1,2$ is understood, and the string tension $g$ is related to the `t Hooft coupling $\lambda$ as $g={\sqrt\lambda\ov 2\pi}$. 

The last term in eq.(\ref{YforQ}) shows that for small values of $h$, the functions $Y_\pm^{(\a)}$ should have an expansion of the form
\bea
Y_\pm^{(\a)} = 1 + h A_\pm^{(\a)} +\cdots\,.
\eea
Then the last term in eq.(\ref{YforQ}) obviously behaves as $\log h$ for small $h$, and we get
 \bea
 - \log Y_Q = - 2\log h \star K_{yQ} + {\rm finite\ terms}\,.
 \eea
 Taking into account that  $1 \star K_{yQ} = 1$, we conclude that $Y_Q$ behaves as $h^2$ 
 \bea
 Y_Q = h^2 B_Q +\cdots\,,
 \eea
 and, therefore, 
 the ground state energy has the following small $h$ expansion
\begin{equation}
E_h (L) = -h^2  \int {du \over 2\pi} \, \sum_{Q=1}^\infty {d\tp^Q \over du} \,  B_Q  + \cO (h^3).
\label{energyL in h}
\end{equation}
Expanding all the Y-functions  around the na\"ive solution \eqref{infinite sol} \begin{alignat}{3}
Y_Q &\approx h^2 B_Q \,,&\qquad Y_\pm^{(\a)} &\approx 1 + h A_\pm^{(\a)} + h^2 B_\pm^{(\a)} \,,\notag \\[1mm]
Y_{M|vw}^{(\a)} &\approx A_M^{(\a)} + h B_{M|vw}^{(\a)} \,,&\qquad Y_{M|w}^{(\a)} &\approx A_M^{(\a)} + h B_{M|w}^{(\a)}\,,
\label{Y expansion in h}
\end{alignat}
one can derive equations for the coefficients $A$'s and $B$'s by substituting the expansions into the TBA equations. 

 It turns out that the following conditions are consistent with the series expansion of the TBA equations up to the first order in $h$
\begin{equation}
B_{M|w}^{(\a)} = B_{M|vw}^{(\a)} \quad \Leftrightarrow \quad A_-^{(a)} = A_+^{(\a)} = 0.
\label{ansatz for vww}
\end{equation}
Then, the TBA equations for $B_Q$ ($Q$-particles), and $A_M^{(\a)}$ ($w$-strings) close within themselves, and take the following simple form\footnote{We use the simplified equations for $vw$- and $w$-strings from appendix 6.3 of \cite{AF09b}.}
\begin{alignat}{3}
- \log B_Q &= L \, \tH_Q - \log ( 1 + \frac{1}{A_M^{(\a)}} ) \star K^{MQ}_{vwx} \,,
\label{A for Q ansatz} \\[1mm]
\log A_M^{(\a)} &=  \log (1 + A_{M-1}^{(\a)})(1 + A_{M+1}^{(\a)})  \star s\,.
\label{eq for AM ansatz}
\end{alignat}
These equations admit a solution with all $A_M^{(\a)}$ being constants independent of $u$. In this case, taking into account that $1\star s = {1\ov 2}$, the second equation reduces to the following form
\begin{equation}
( A_M^{(\a)})^2 =(1 + A_{M-1}^{(\a)})(1 + A_{M+1}^{(\a)}) \quad {\rm for} \ \ M \ge 1, \qquad A_0^{(\a)} = 0.
\label{eq for AM const}
\end{equation}
It has the following regular solution\footnote{One can also check that 
eq.(\ref{AM constant ansatz}) solves 
the original TBA equations for $vw$- and $w$-strings, by using $n_{xv}^{QM} \equiv 1 \star K_{xv}^{QM} = M-1$ for $M-1<Q$ and $n_{xv}^{QM} = Q$ for $M-1 \ge Q$, together with $n_{MM'} \equiv 1 \star K_{MM'} = 2 M$ for $M < M'$ and $n_{MM'} = 2 M' - \delta_{M M'}$ for $M \ge M'$.} 
\begin{equation}
A_{M-1}^{(\a)} = M^2-1 \qquad( M \ge 1),
\label{AM constant ansatz}
\end{equation}
which coincides with 
 the constant solution of a Y-system 
 %corresponding to the ground state energy 
 discussed in \cite{BH03a,BH03b}.
 
We can now find $B_Q$ from eq.(\ref{A for Q ansatz}).  For constant 
$A_M^{(\a)}$ the convolution terms in \eqref{A for Q ansatz} can be computed by using 
that $1\star K^{MQ}_{vwx} = n_{vwx}^{M,Q}$, where
the integers $n_{vwx}^{M,Q}$  satisfy 
$n_{vwx}^{M,Q} = M-1$ for $M < Q-1$, and 
$n_{vwx}^{M,Q} = Q$ for $M \ge Q-1$.
A simple computation then gives
\begin{alignat}{3}
\log ( 1 + \frac{1}{A_M^{(\a)}} ) \star K^{MQ}_{vwx} &= \sum_{M=2}^{Q-1} M \log ( 1 + \frac{1}{M^2-1} ) + Q \sum_{M=Q}^\infty \log ( 1 + \frac{1}{M^2-1} ),
\notag \\
&= \log ( \frac{2 \ssp (Q-1)^Q}{Q^{Q-1}} ) + Q \log ( \frac{Q}{Q-1} ) = \log 2 \ssp Q,
\label{AM constant sum}
\end{alignat}
for each $\alpha=1,2$, and therefore 
\bea\la{YQ}
Y_Q = 4 \ssp h^2 \ssp Q^2 \, e^{-L \tH_Q}+ \cO (h^3)\,.
\eea 
Taking into account that  the  energy of a mirror $Q$-particle can be written in the form
\bea
 \tH_Q = \log {x^{Q-}\ov x^{Q+}}\,, \qquad x^{Q\pm}(u)=x(u\pm{i\ov g} \, Q)\,, 
\eea
the $Y_Q$-functions acquire the form
\bea\la{YQ2}
Y_Q = 4 \ssp h^2 \ssp Q^2 \, \left({x^{Q+}\ov x^{Q-}}\right)^L+ \cO (h^3)\,.
\eea 
Since  the variables $x^{Q\pm}$ are expressed in terms of the function $x(u)={1\ov 2}\big(u -i \sqrt{4-u^2}\big)$, the $Y_Q$-functions are not analytic on the $u$-plane for any value of $L$, and have there two cuts.
On the other hand it is known that the dispersion relation for $Q$-particles is uniformized 
 in terms of the $z$-torus rapidity variable \cite{Janik}, 
 and 
 the ratio $x^{Q+}/x^{Q-}$ is given by $(\cn\, z + i \sn\, z)^2$, which is real when $z$ is on the real axis of mirror region.\footnote{The elliptic modulus $k=-4g^2/Q^2$ of the Jacobi functions depends on  $Q$, and, therefore, the periods of the torus depend on $Q$ too.}
 We conclude, therefore, that the $Y_Q$-functions are meromorphic on 
the $z$-torus if 
\begin{equation}
L = {1\ov T} \quad \text{is integer or half-integer.}
\end{equation}
We do not expect $Y_Q$ to be analytic on the $z$-torus for finite values of $h$ because then the dressing factor \cite{AFS}-\cite{BES} would start contribute to the equations for $Y_Q$, and it is known that the dressing factor has infinitely many cuts on the $z$-torus. Nevertheless, as was recently shown in \cite{AF09c}, the dressing factor\footnote{The BES dressing factor \cite{BES}  was proven \cite{Volin} to be the minimal solution of crossing equations \cite{Janik}.} is holomorphic in the union of the physical regions of the string and mirror models, and, therefore, it is natural to require the $Y_Q$-functions to be meromorphic there too. 

As we have shown above, an unusual consequence of this requirement is that the circumference of the circle the light-cone string theory lives on, and the temperature of the mirror theory are quantized. Let us also mention that the charge quantization is the first step to understand the full $\psu(2,2|4)$ symmetry of the string spectrum from the TBA approach.\footnote{Note that the $\psu(2|2)^2$ charges of the string theory and the mirror theory are not physically equivalent due to double Wick rotation.}.

Finally, the ground state energy at the leading order in $h$ and arbitrary $L$ is  given by
\begin{equation}
E_h (L) \approx - h^2 \int {du \over 2\pi} \, \sum_{Q=1}^\infty {d\tp^Q \over du} \, 4 \ssp Q^2 \, e^{-L \tH_Q} = - h^2 \sum_{Q=1}^\infty \int {d\tp^Q \over 2\pi} \,   4 \ssp Q^2 \, e^{-L \tH_Q}\,.
\label{energyL constant ansatz}
\end{equation}
By using eq.(\ref{def:mirror energy}) for the energy of a $Q$-particle, one can show that the sum is convergent for $L>2$. For $L=2$ the series in $Q$ diverges as ${1\ov Q}$.  We do not understand the reason for the divergency. Note that $L=2$ is the lowest value the total light-cone momentum can have. 

%%%%%%%%%%%%%%%%%%%%%%%%%
\subsection{General ${h}$ at large ${L}$}

It is also of interest to consider the large $L$ asymptotics of the ground state energy with $h$ fixed. 
In this case we expect that the finite-size corrections to the energy of the ground state can be also computed  by introducing a twist in the generalized L\"uscher formula \cite{Luscher85, JL07, BJ08, HS08b}:
\begin{equation}
E_{\rm gL} (L) = - \int {du \over 2\pi} \, \sum_{Q=1}^\infty {d\tp^Q \over du} \, e^{-L \, \tH_{Q}} \; {\rm tr}_Q e^{i (\pi + h) F}  + \cO ( e^{-2L \, \tH_{Q}} )\,.
\label{JL gnd formula}
\end{equation}
Here  the trace runs through all $16 \ssp Q^2$ polarizations of a $Q$-particle state, and $F$ is the fermion number operator which in our case is equal to the difference 
$F_1-F_2$ where $F_\a$ is equal to the 
 number 
of $y^{(\a)}$-particles. 

Computing the trace in \eqref{JL gnd formula}
\begin{equation}
{\rm tr}_Q \, e^{i (\pi + h) F} \equiv {\rm tr}_Q \, e^{i (\pi + h) (F_1 - F_2)} = 2 Q ( 1 - e^{i h} ) \cdot 2 Q ( 1 - e^{-i h} ) \,,
\label{twisted trace}
\end{equation}
and substituting the result back into \eqref{JL gnd formula}, we obtain
\begin{equation}
E_{\rm gL} (L) = - \int {du \over 2\pi} \, \sum_{Q=1}^\infty {d\tp^Q \over du} \, 16 \ssp Q^2 \sin^2 \frac{h}{2} \; e^{-L \, \tH_{Q}} + \cO ( e^{-2L \, \tH_{Q}} )\,.
\label{JL gnd general h}
\end{equation}
At small values of $h$ the formula obviously agrees with (\ref{energyL constant ansatz}).
We will see in a moment that it matches precisely  the large $L$ asymptotics of the ground state energy with $h$ fixed computed by using the TBA equations.

The series expansion of Y-functions in terms of $e^{-L \, \tH_Q}$ can be performed almost in the same manner as the small $h$ expansion. We can find a consistent solution at the leading order assuming the following expansion of the Y-functions
\begin{equation}
Y_Q \approx B_Q \; e^{-L \, \tH_Q} \,,\quad Y_\pm^{(\a)} \approx A_\pm^{(\a)},\quad Y_{M|w}^{(\a)} \approx A_{M|w}^{(\a)}\,, \quad Y_{M|vw}^{(\a)} \approx A_{M|vw}^{(\a)} \,,
\label{Ys general h}
\end{equation}
where $B_Q$ and $A$'s are independent of $u$.

Then, $A_{M|w}^{(\a)}= A_{M|vw}^{(\a)}$ given by the same constant ansatz of \eqref{AM constant ansatz} solve the equations for $vw$- and $w$-strings if  $A_\pm^{(\a)}=1$. 
The TBA equation for $Q$-particles \eqref{YforQ} then takes the same form as for the small $h$ case with the only change $h^2\to 4\sin^2{h\ov 2}$, and, therefore, its solution is given by  
$$B_Q = 16 \ssp Q^2 \sin^2 {h\ov 2}\,.$$ 
Thus, the energy of the ground state becomes
\begin{equation}
E_h (L) = - \int {du \over 2\pi} \, \sum_{Q=1}^\infty {d\tp^Q \over du} \, 16 \ssp Q^2 \, \sin^2 \frac{h}{2} \; e^{-L \, \tH_Q} + \cO ( e^{-2L \, \tH_Q} ),
\label{energyL general h}
\end{equation}
which 
%is consistent with \eqref{energyL constant ansatz} for $h$ small, and 
 completely agrees with the generalized L\"uscher formula (\ref{JL gnd general h}). 
 It is worth stressing that the contribution of the $vw$-strings  is crucial for the agreement.\footnote{There is simple generalization of this agreement. If we generalize the solution \eqref{eq for AM const} to $A_{M-1} = \sin^2 (Mz/2)/\sin^2 (z/2) - 1$, the $Y_Q$-functions agree with a so-called elliptic genus, ${\rm tr}_Q e^{- L \tH_Q + i (\pi + h)F + i z J_3}$ with $J_3=\sigma_3 \cdot \bb{L} + \sigma_3 \cdot \bb{R}$ in the notation of \cite{AF08a}.}

The energy, as well as the generalized L\"uscher formula, obviously diverges again logarithmically for $L=2$.\footnote{Eq.(\ref{energyL general h}) was derived for large $L$. It  nevertheless  is valid for finite $L$ and small $g$ because in this case $Y_Q$ is decreasing as $g^{2L}$.} Note that the corresponding formula in the computation of the anomalous dimension of the Konishi operator \cite{BJ08} contains additional factor of $e^{-2 \tH_Q}$ coming from the transfer matrix, which renders the series convergent.

Let us finally mention that for $h=\pi$ the formula \eqref{energyL general h} should give the energy of the non-BPS ground state of the light-cone string theory in the sector with anti-periodic fermions, and through the AdS/CFT correspondence the scaling dimension 
of the dual ${\cal N}=4$ SYM operator. It would be interesting to identify this operator and compute its perturbative scaling dimension.

%%%%%%%%%%%%%%%%%%%%%%%%%
\subsection{Analyticity of Y-system}
We recall  that the simplified TBA equation for $Y_1$-function takes the form \cite{AF09b}
\bea
\la{YforQ1}
\log Y_{1}&=&
\log{\left(1-{e^{ih_1} \ov Y_{-}^{(1)}} \right)\left(1-{e^{ih_2} \ov Y_{-}^{(2)}}\right)\ov 1 +  {1\ov Y_{2}} }\star s - \Delta\star s \,,~~~~~ \eea 
where 
\bea\la{addit}
\Delta&=&\log\left(1-{e^{ih_1} \ov Y_{-}^{(1)}}
\right)\left(1-{e^{ih_2} \ov Y_{-}^{(2)}}\right) \big(\theta(-u-2)
+\theta(u-2)\big)
\\\nonumber
&+& L\, \check{\cal E} -\log\left(1-{e^{ih_1} \ov Y_{-}^{(1)}}
\right)\left(1-{e^{ih_2} \ov Y_{-}^{(2)}}\right) \left(1-{e^{ih_1}
\ov Y_{+}^{(1)}} \right)\left(1-{e^{ih_2} \ov Y_{+}^{(2)}}\right)
\star \check{K}
\\
\nonumber&-& \log\left(1 +  {1\ov Y_{M|vw}^{(1)}} \right)\left(1 +
{1\ov Y_{M|vw}^{(2)}} \right)\star \check{K}_M +2\log\left(1+Y_{Q}
\right)\star {\check K}^\Sigma_{Q}  \,,~~~~~ \eea 
is  the obstruction to have the Y-system outside $u \in [-2,2]$.
 It has been shown in \cite{AF09b} that the TBA equations may lead to a usual Y-system only if $\Delta$ vanishes on any solution. Let us also stress that the vanishing of $\Delta$ is only one of the several necessary conditions the Y-functions should satisfy, see \cite{AF09b} for a detailed discussion. These conditions are trivially satisfied at the leading order in the small $h$ expansion, but it is unclear how to verify all the necessary conditions for finite $h$.
 
To compute $\Delta$ for the ground state solution we use that 
$1 \star \check{K} ={1\ov 2} \big(\theta(-u-2)
+\theta(u-2)\big)$, and $1\star \check{K}_M=0$, and in the both small $h$ and large $L$ cases we get the following leading term 
\bea\la{Dgr}
\Delta =  L\, \check{\cal E} = L \,\log  \frac{x (u + i0)}{x (u - i0)}  \neq 0 \qquad {\rm for} \ \ u \in (-\infty,-2) \cup (2,\infty)\,.
\eea
Since $\Delta$ does not vanish, 
the Y-functions are not analytic in the complex 
$u$-plane,\footnote{We have seen this already from the explicit solution (\ref{YQ2}).} and
the 
TBA equations do not lead to an analytic Y-system.  
It still might be possible to define (but not to derive) the Y-system on an infinite genus Riemann surface,\footnote{We thank Pedro Vieira for a discussion of this possibility.} and require the validity of the Y-system equations on its particular sheet.  

To see how this might work, let us recall that the Y-equation is obtained from \eqref{YforQ1} by applying to it an operator $s^{-1}$, and has the following form \cite{AF09b}
\bea
\la{YforQ1b}
e^{\Delta(u)}Y_{1}(u+{i\ov g}-i0)Y_{1}(u-{i\ov g}+i0)=
{\big(1-{e^{ih_1} \ov Y_{-}^{(1)}} \big)\big(1-{e^{ih_2} \ov Y_{-}^{(2)}}\big)\ov 1 +  {1\ov Y_{2}} }\,.~~~~~ \eea 
The explicit ground state solution \eqref{YQ2} and \eqref{Dgr} then show that 
the jump discontinuity of $\log Y_{1}(u\pm{i\ov g})$ across the real $u$-line is given by 
$\pm\Delta(u)$, and, therefore, for the ground state solution \eqref{YforQ1b} can be written in the form
\bea
\la{YforQ1c}
Y_{1}(u+{i\ov g}\pm i0)Y_{1}(u-{i\ov g}\pm i0)=
{\big(1-{e^{ih_1} \ov Y_{-}^{(1)}} \big)\big(1-{e^{ih_2} \ov Y_{-}^{(2)}}\big)\ov 1 +  {1\ov Y_{2}} }\,.~~~~~ \eea 
Thus, we conclude that the Y-system equation for $Y_1$ might hold on the $u$-plane with the cuts running from $\pm 2\pm {i\ov g}$ to infinity along the horizontal lines if the shifts 
upward and downward are defined with the infinitesimal parts of the same sign. 
The equations for $Y_Q\ (Q \ge 2)$ would then induce infinitely many cuts on the $u$-plane with the branch points located at $\pm 2\pm {i\ov g}Q$. 

Let us stress again that the Y-system equations can have the canonical form only on a particular sheet of the infinite genus Riemann surface, and probably would take different forms on other sheets. 
In this respect it is similar to the \AdS crossing equations \cite{Janik}. It would be interesting (and necessary) to understand the corresponding transformation properties of the Y-system.
This is in contrast to relativistic models, and 
it is unclear to us if such a Y-system would be useful for analyzing the spectrum along the lines of  \cite{Fendley:1997ys}-\cite{Heg},\cite{GKV08}. 

%%%%%%%%%%%%%%%%%%%%%%%%%%%%%%%%%%%%%%%

\section{Conclusions}

In this paper we have analyzed the TBA equations for the ground state energy of the light-cone superstring on the \AdS background. We have shown that the natural condition of the analyticity of the solution of the TBA equations on the union of the physical regions of the string and mirror  \AdS  models leads to the quantization of the circumference of the cylinder the string theory lives on, or, equivalently, to the temperature quantization of the mirror model.  

The temperature quantization is a new phenomenon never seen before, and it is the simplest manifestation of the $\psu(2,2|4)$ symmetry of the superstring and ${\cal N}=4$ SYM spectrum. The full  $\psu(2,2|4)$ symmetry can been seen only in the TBA equations for excited states, and it is of utmost importance to prove it.

We also have analyzed the TBA equations for large $L$ and  finite $h$, and have shown that the ground state energy completely agrees with the twisted L\"uscher formula.
We have observed that the energy is logarithmically divergent for $L=2$. It would be interesting to understand the origin of the divergency.

The TBA equations describe the spectrum of light-cone superstring on \AdS only for $h=0$ or $h=\pi$. The general $h$ case should describe something which goes beyond the usual correspondence between \AdS\ superstring and $\cN=4$ super Yang-Mills.

On string theory side, one can introduce magnetic flux coupled to worldsheet fermions. A fermion acquires an extra phase $\chi \to e^{ih} \chi$ when it goes around the worldsheet cylinder. This magnetic flux is topological (or Aharanov-Bohm type) in the sense that the phase of a worldsheet fermion remains unchanged for any contractible cycle on the cylindrical worldsheet. The magnetic field must be a spacetime singlet, because the extra phase does not change the spacetime index of fermions. The magnetic field can couple to worldsheet fermions by replacing $\partial_a \chi$ with $(\partial_a - i A_a) \chi$ in the light-cone superstring sigma model.

At weak coupling, the TBA energy is a quantity of order $\cO (g^{2L})$ for any length $L$ operator and for any $h$. Since the effect of non-zero $h$ can be observed only beyond the wrapping order, the chemical potential does not modify the  local structure of the dilatation operator of $\cN =4$ super Yang-Mills. Thus, on the gauge theory side, we expect that there is a way to interpret the chemical potential as a modification of local operators, e.g. as in orbifold models, see \cite{BR},  rather than a deformation of the Lagrangian. 
A convincing interpretation is not known to us, and it would be interesting to clarify this.

%%%%%%%%%%%%%%%%%%%%%%%%%%%%%%%%%%%%%%%
\section*{Acknowledgements}
We are grateful to Gleb Arutyunov, Yasuyuki Hatsuda, Matthias Staudacher and Pedro Vieira for valuable discussions. This project was finished while S.F. was visiting Max-Planck-Institut f\"ur Gravitationsphysik
Albert-Einstein-Institut. 
The work of S.F. and R.S. was supported by the Science Foundation Ireland under Grant
No. 07/RFP/PHYF104.  The work of S.F. 
was
supported in part by  a
one-month Max-Planck-Institut f\"ur Gravitationsphysik
Albert-Einstein-Institut grant. 

%%%%%%%%%%%%%%%%%%%%%%%%%%%%%%%%%%%%%%%

\end{document}